# Implementation of an AI-based MRD evaluation and prediction model for multiple myeloma

**Jianfeng Chen [1], Jize Xiong[2], Yixu Wang[3], Qi Xin[4], Hong Zhou[5]**

[1]Statistics, Kansas State University, Manhattan, Kansas, USA
[2] Computer Information Technology, Northern Arizona University, Flagstaff AZ, USA
[3] Computer Science, Independent Researcher, Beijing, China
[4] Management Information Systems, University of Pittsburgh, Pittsburgh, PA, USA
[5] Computer Technology, Peking University, Beijing, China

\* **Corresponding author**: HongZhou (Email: ylou13@hawk.iit.edu)

**Abstract:** *With the application of hematopoietic stem cell transplantation and new drugs, the progression-free survival rate and overall survival rate of multiple myeloma have been greatly improved, but it is still considered as a kind of disease that cannot be completely cured. Many patients have disease recurrence after complete remission, which is rooted in the presence of minimal residual disease MRD in patients. Studies have shown that positive MRD is an independent adverse prognostic factor affecting survival, so MRD detection is an important indicator to judge the prognosis of patients and guide clinical treatment. At present, multipa-rameter flow cytometry (MFC), polymerase chain reaction (PCR), positron emission tomography (positron emission) Several techniques, such as PET/computer tomography (CT), have been used for MRD detection of multiple myeloma.However, there is still no cure for the disease. "IFM2013-04" four clinical studies confirmed for the first time that proteasome inhibitors (PIs) and immunomodulatory drugs, The synergism and importance of the combination of IMiDs in the treatment of MM, the large Phase 3 clinical study SWOG SO777 compared the combination of bortezomib plus lenalidomide and dexamethasone. The efficacy of VRD and D established the status of VRD first-line treatment of MM, and due to the good efficacy of CD38 monoclonal antibody in large clinical studies, combination therapy with VRD has been recommended as the first-line treatment of MM. However, to explore the clinical value and problems of applying artificial intelligence bone marrow cell recognition system Morphogo in the detection of multiple myeloma minimal residual disease (MRD)*

## 1. INTRODUCTION

Rapid pharmacological and technological developments in recent years can greatly reduce early mortality and improve survival rates in MM patients, significantly improving the prognosis of MM patients. Therefore, early identification of patients with MM and accurate risk stratification and prognosis assessment are essential for personalized treatment of MM. Existing MM risk stratification systems include the DS staging system, the revised international staging system (R-ISS), and the Mayo myeloma stratification and risk-adjusted Treatment Stratification system. These staging systems combine serological and genetic factors reflecting disease burden to stratify MM patients to assess their prognosis. However, these existing markers have problems of inadequate systemic representation and low sensitivity, and more convenient, sensitive and specific biological markers need to be combined to stratify patients more accurately[1-2].

Therefore, under the background of the strong combination and development of artificial intelligence and medical industry, the use of new data algorithms to detect some special causes has become the biggest concern of the medical community, so the application of MRD predictive model can achieve the rapid detection and diagnosis of multiple myeloma to a large extent. Relevant studies have shown that positive MRD is an independent adverse prognostic factor affecting survival[3].Therefore, MRD detection is an important indicator to judge the prognosis of patients and guide clinical treatment. At present, multiparameter flow cytometry (MFC), polymerase chain reaction (PCR), positron emission tomography (positron emission) tomography (PET)/ computer tomography (CT) and other techniques have been used in MRD detection of multiple myeloma. However, the above technologies have the characteristics of high cost and long time. Cell morphology is a mature test technology, plasma cells can be clearly identified through microscope observation, but the traditional cell morphology through manual classification, the number of classified cells is small, plasma cell recognition rate depends on the professional quality of the examiner and other defects, which makes the traditional cell morphology cannot be applied to multiple myeloma MRD detection. artificial intelligence (AI) is a new computer technology that is currently flourishing, and its branches include convolutional neural net-work (CNN) and fuzzy expert systems (fuzzy expert systems) Technologies such as system FES, evolutionary computing, and hybrid intelligent systems have all been used in different clinical Settings, and CNN technology in particular has been applied to the clinical diagnosis of radiology and pathology[4-5]. At present, many institutions have studied the application of artificial intelligence based on CNN to the cell morphological diagnosis of blood diseases. Artificial intelligence convolutional neural network (AI-CNN) cell morphology combined with artificial intelligence cell recognition technology,Is it used in clinical diagnosis of MRD multiple myeloma? This is the focus of this study.

## 2. RELATED WORK

**2.1 MRD techniques for assessing prognosis in MM**

At present, the detection of MM-MRD has reached a certain consensus at home and abroad 12%, usually using flow cytometry or gene sequencing to identify residual tumor cells in the bone marrow, and using PET/CT imaging technology to detect whether there is residual disease outside the bone marrow. Bone marrow cell morphology is currently the most widely used method for evaluating tumor cell load. Because traditional bone marrow cell morphology examination uses manual classification and counting of 200-500 cells, its sensitivity is low and it is not suitable for MRD detection[6]. However, with the rise of AI image recognition technology, the problem of cell morphology inspection about the order of magnitude is expected to be solved to a certain extent CNN algorithm CNN is a kind of deep learning, and its advantage is that it can directly input images into the neural net end to avoid the complex feature extraction and data reconstruction process in the traditional algorithm, and it has certain invariance to the translation scale scaling and other forms of deformation of two-dimensional images[7].

MRI has a variety of imaging sequences and imaging parameters, and a variety of MRD techniques have been used to evaluate the degree of bone marrow involvement in MM.

A meta-data analysis looked at 16 published studies, including 2,076 patients with ALL, and quantified the relationship between survival outcomes and MRD status. Negative MRD was found to be associated with event-free survival (EFS) and overall survival (OS). Evidence for the prognostic value of MRD testing was consistent across all disease subtypes, age, treatment received, stage of treatment at the time of testing, and MRD testing methods.

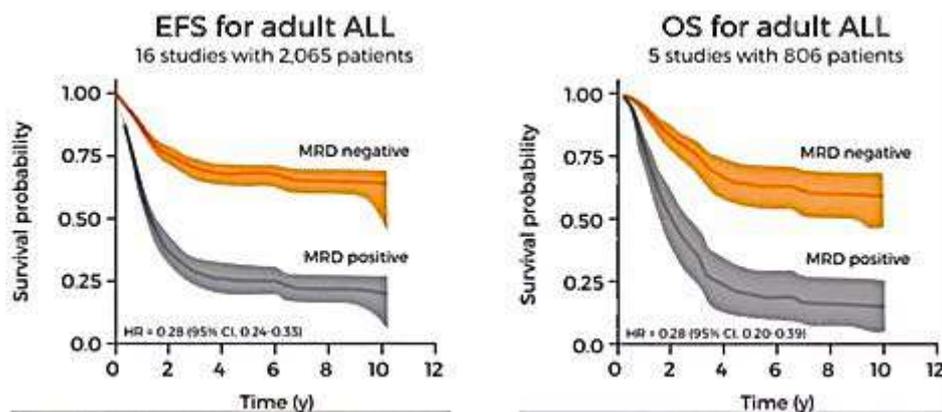

**Figure 2:** EFS and OS multiple myeloma recurrence diagram

These techniques include conventional MRD sequences such as T1-weighted imaging, T2-weighted and lipid-pressure imaging, and TTI-weighted random-enhanced imaging, as well as dynamic contrast-enhanced MRI. MM lesions typically showed low signal in T1-weighted images, high signal in T2-weighted images, and often enhanced signal in vanadium enhanced images. On the WB-DWI images, the diffusion of MM lesions was limited and showed a high signal. The apparent diffusion coefficient (ADC) could be used to quantify the moisture diffusion. The m DIXON technique can be used to quantify bone marrow fat content by fat fraction (FF). IVIM technology can quantify the diffusion and perfusion of bone marrow[8]

**2.2 Application of different MRI sequences in MM-MRD prognosis**

The conventional MRI sequences used to evaluate MM bone marrow infiltration were T1 weighted,T2 weighted lipid-enhanced, and TI weighted lipid-enhanced. These sequences can clearly show MM bone marrow infiltration pattern and focal number[9]. MM bone marrow infiltration pattern can be divided into 5 types: diffuse infiltration type, focal type, diffuse focal type, salt-pepper type and normal type. The pattern of bone marrow infiltration is related to the stage of DS. Studies have shown that the normal type and the salt-pepper type are in the stage of DS I, most of the diffuse type are in the stage of I, and most of the focal type and diffuse focal type are in the stage of I or I. The pattern of bone infiltration is related to the stage of ISS, and studies have shown that patients with diffuse infiltration are mostly located in ISS stage I.

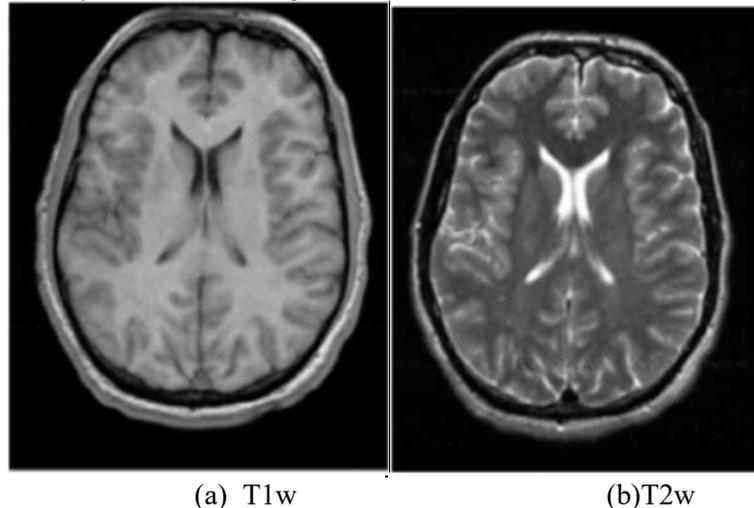

(a) T1w          (b)T2w
**Figure 1:** T1w is the reference image and T2w is the input image

T1-weighted imaging (T1WI) ---- highlights differences in tissue T1 relaxation (longitudinal relaxation)
T2-weighted imaging (T2WI) ---- highlights differences in tissue T2 relaxation (transverse relaxation).
On any sequential image, the larger the transverse magnetization vector at the time of signal acquisition, the stronger the MR Signal.
T1 weighted image short TR, short TE -- T1 weighted image, T1 image characteristics: the shorter the tissue T1, the faster the recovery, the stronger the signal; The longer the T1 of the tissue, the slower the recovery and the weaker the signal.
T2 weighted image length TR, long TE -- T2 weighted image, T2 image characteristics: the longer the tissue T2, the slower the recovery, the stronger the signal; The shorter the T2 of the tissue, the faster the recovery, and the weaker the signal. Proton density weighted image long TR, short TE - proton density weighted image, image features: the larger the rH of the tissue, the stronger the signal; The smaller the rH, the weaker the signal.
It is generally believed that the high signal on the T1-weighted image is mostly caused by bleeding or adipose tissue. However, recent studies have shown that T1-weighted high signal can still be seen in a variety of intracranial lesions, including tumors, cerebrovascular diseases, metabolic diseases and some normal physiological states. Under the excitation of radio frequency pulse, the energy absorbed by hydrogen protons in human tissues is in an excited state. In the relaxation process, the hydrogen proton releases its absorbed energy into the surrounding environment. If the proton and the proton in the lattice also precess at a frequency similar to the Larmor frequency, then the energy release of the hydrogen proton is faster.
The shorter the T1 relaxation time of the tissue, the higher the signal intensity of the T1-weighted image. There are three conditions in which T1 relaxation time is shortened: one is the binding water effect; The other is paramagnetic substance; The third is a lipid molecule,There is a general consensus that the prognosis of patients with diffuse infiltration is worse than that of other infiltration types, but the prognosis of focal type is significantly different. Some studies have found that the more lesions detected by MRI, the higher the degree of bone destruction, and the focal lesions beyond the cortical bone boundary significantly affect the survival rate[10-13]. The presence of local infiltrating lesion on MRI is a poor prognostic factor for asymptomatic MM progression to symptomatic MM. In symptomatic MM, patients with more than 7 MRI sites had a poor prognosis, while patients with less than or equal to 7 sites had a similar survival rate. (Routine MRI sequences, in addition to showing MM infiltration pattern and number of sites, can also semi-quantify the degree of bone marrow infiltration.)

## 3. METHODOLOGY

In this study, the consistency between Morphogo and cell morphology in plasma cell recognition was also good. In AI-MRD VS.M-MRD, the Kappa value of Kappa induction test continued to increase with the increase of the number of recognized cells, and Kappa=0.667(P=0.001) in the 2000 group, indicating that the results of AI-MRD and M-MRD were highly consistent when the recognition level of AI cells reached 2000. In further data analysis, the sensitivity and accuracy of MRD results between AI and cell morphology also increased significantly with the increase of the number of cells recognized by AI, and the specificity was always maintained at the level of 90% to 100%. This shows that AI can achieve almost complete agreement with the results of artificial cell morphological examination when determining negative results of patients, and the consistency increases with the increase of the number of recognized cells when determining positive results.

### 3.1 Model Architecture

MorphogoMM-3 is an ongoing multi-center, open-label, single-arm, phase 2 study investigating the efficacy and safety of eranata mab in patients with relapsed or refractory multiple myeloma. Eligible patients were 18 years of age and older, had previously been diagnosed with multiple myeloma and met the International Myeloma Working Group (IMWG) criteria for measurable disease, had good bone marrow function (platelets $\geq 25\times 10^9$ l$^{-1}$, absolute neutrophilic counts $\geq 1.0\times 10^9$ l$^{-1}$, and had good bone marrow function (platelet $\geq 25\times 10^9$ L$^{-1}$). Hemoglobin $\geq 8$ g dl$^{-1}$), good liver function (total bilirubin $\leq 2$ times the upper limit of normal ($\leq 3$ times ULN if Gilbert syndrome is demonstrated), aspartate aminotransferase $\leq 2.5$ times ULN, Alanine aminotransferase $\leq 2.5$ times ULN) and renal function (creatinine clearance $\geq 30$ ml min$^{-1}$), and Eastern Cooperative Oncology tissue (ECOG) performance status $\leq 2$. Patients must be resistant to at least one proteasomal inhibitor, one immunomodulatory drug, and one anti-CD38 antibody, and be relapsed or refractory to the last anti-myeloma regimen. Patients in group A could not receive targeted BCMA therapy. From February 9, 2021 to January 7, 2022, A total of 123 patients were enrolled in the Group A cohort and received doses at 47 study sites in ten countries.

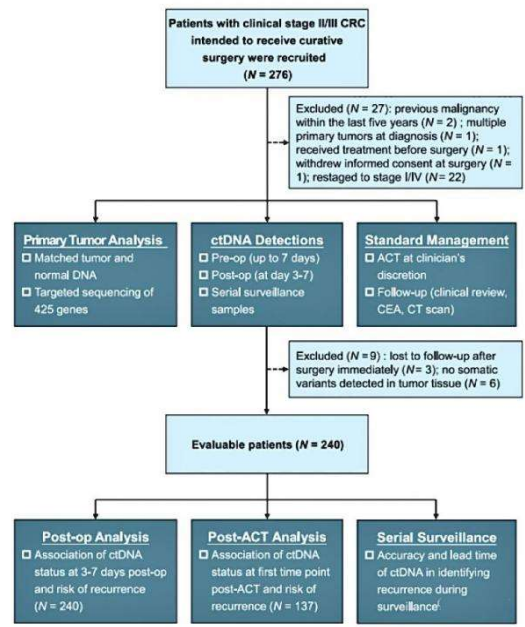

**Figure 3:** Morphogo MM-MRD-3 evaluation model

### 3.2 An antibacterial agent to prevent infection

Peripheral neuropathy, defined as motor dysfunction and sensory neuropathy, was reported in 17.1% and 13.8% of patients, respectively. Of these patients, 14.1% and 35.3% had a history of motor dysfunction and sensory neuropathy, respectively. The most common ($\geq 5$%) neuropathic events were muscle spasm and peripheral sensory neuropathy (7.3% each). Motor dysfunction and sensory neuropathy occurred in 1 (0.8%) and 0 cases, respectively, with no grade 4 or 5 events. Among 119 patients who received a two-step priming dose regimen, 56.3% developed cytokine release syndrome (CRS). All CRS incidents were Level 1 (42.0%) or Level 2 (14.3%), and no level 3 or above incidents were reported. With respect to the most recent dose, the median time to onset of CRS was 2.0 days (range: 1.0-9.0 days) and the median time to resolution was 2.0 days (range: 1.0-19.0 days). Overall, 98.8% of CRS events occurred in the first three doses, and 90.6% occurred in the incremental dose. One patient (0.8%) developed a Grade 1 CRS event after the fourth or later dose of elranatamab (Figure 4). Eighteen (15.1%) patients had more than one CRS event. Tocilizumab and corticosteroids were used in 22.7% and 8.4% of patients with CRS, respectively. ICANS occurred in 4 of 119 patients (3.4%), and all events were grade 1 or 2. Supportive treatment for ICANS included corticosteroids (1.7%), tocilizumab (1.7%), and levetiracetam for seizure prevention (0.8%). None of the patients permanently discontinued elranatamab due to the development of CRS or ICANS.

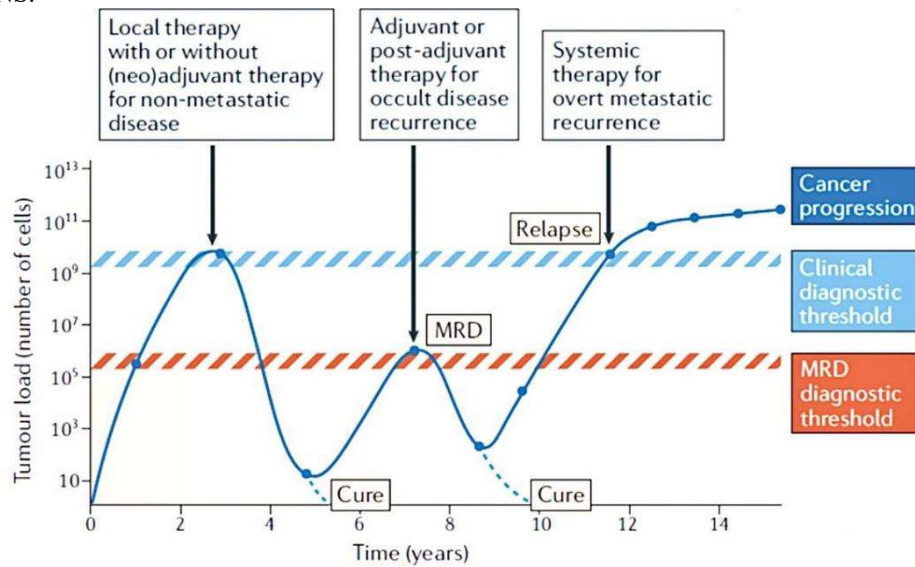

**Figure 4:** Prediction of recurrence of multiple myeloma in patients

In this study of patients with relapsed or refractory multiple myeloma, the MM-MRD predictive diagnostic model showed a high rate of deep and long-lasting responses, including in patients achieving ≥CR, with a manageable safety profile. Dosing with two incremental initiation dosing regimens successfully reduced the incidence and severity of CRS, a predictable scenario that supports the potential for outpatient dosing. Despite the need for additional follow-up, MM-MRD was observed to maintain or deepen responses after switching to a biweekly schedule. And biweekly dosing can provide greater convenience for patients while potentially being less toxic.

## 4. CONCLUSION

At present, the detection of MM-MRD has reached a certain consensus at home and abroad. Flow cytometry or gene sequencing is usually used to identify residual tumor cells in the bone marrow, and PET/CT imaging technology is used to detect whether there is residual disease outside the bone marrow. Bone marrow cell morphology is currently the most widely used method to assess tumor cell load, because traditional bone marrow cell morphology examination uses manual classification and counting of 200 to 500 cells. Its low sensitivity is not suitable for MRD detection, but with the rise of AI image recognition technology, the problem of cell morphology inspection about the order of magnitude is expected to be solved to a certain extent. According to the results of the study, it was also found that when patients obtained CR before harvesting, the possibility of their samples being contaminated by tumor cells was significantly reduced[14-15]. However, for patients who obtained VGPR or PR before harvesting, there was no statistical significance in the positive proportion of samples being contaminated by tumor cells between the two groups, and even if patients obtained CR before harvesting. The level of neoplastic PCs in the SCC was similar to that of patients receiving other therapeutic effects (VGPR and PR), and the difference among the three groups was not statistically significant. In addition, we found that although patients in the CMRD-negative group had a statistically significant deeper pre-harvest remission than those in the CMRD-positive group, this advantage was not maintained after transplantation. Traditional efficacy evaluation (CR/VGPR/PR can not well reflect the MRD condition of patients' bone marrow, even if CR is obtained, clonal PCs may still exist in bone marrow and be mobilized into peripheral blood. Combined with the results of this study, hematopoietic stem cells can be mobilized, dried and transplanted after PR. There was no significant adverse effect on the efficacy of MRD and ASCT[16].

In short, the application of AI for MM-MRD detection has the characteristics of high cell recognition accuracy, fast speed and low cost, but the consistency of AI and flow cytometry results is still not high enough based on morphological cell recognition, but AI still has high growth in cell recognition. In subsequent development, technologies such as cytochemical staining, cellular immunity, molecular biology and full slide imaging technology "5 "should be considered, so as to continuously improve the accuracy of AI in the diagnosis of MM MRD, and provide efficient, accurate and inexpensive testing technologies for clinical use.

In this study, we have introduced a novel methodology tailored for the mechanical domain, specifically focused on predictingthe position of a robot based on sensor data collected from the floor.

## REFERENCES

[1] VAN DE DONK N W CJ,PAWLYN C.YONG K L.Multiple myeloma [].TheLancet,2021,397(10272): 410-27.
[2] MOREAU P HULIN C,MACRO M,et al.VTD is superior to VCD prior tointensive therapy in multiple myeloma: results of the prospective lFM2013-04 trial[J].Blood,2016,127(21): 2569-74.
[3] DURIE B G M,HOERING A, ABIDI M Het al. Bortezomib with lenalidomideand dexamethasone versus lenalidomide and dexamethasone alone in patients withnewly diagnosed myeloma without intent for immediate autologous stem-celltransplant (SWOG S0777): a randomised, open-label, phase 3 trial [J]. The Lancet.2017,389(10068):519-27.
[4] Yimin Ou, Rui Yang, Lufan Ma, Yong Liu, Jiangpeng Yan, Shang Xu, Chengjie Wang, Xiu Li,UniInst: Unique representation for end-to-end instance segmentation,Neurocomputing,Volume 514,2022,Pages 551-562,ISSN 0925-2312.
[5] Chang Che, Bo Liu, Shulin Li, Jiaxin Huang, and Hao Hu. Deep learning for precise robot position prediction in logistics. Journal of Theory and Practice of Engineering Science, 3(10):36–41, 2023.DOI: 10.1021/acs.jctc.3c00031.
[6] Hao Hu, Shulin Li, Jiaxin Huang, Bo Liu, and Change Che. Casting product image data for quality inspection with xception and data augmentation. Journal of Theory and Practice of Engineering Science, 3(10):42–46, 2023. https://doi.org/10.53469/jtpes.2023.03(10).06
[7] Chang Che, Qunwei Lin, Xinyu Zhao, Jiaxin Huang, and Liqiang Yu. 2023. Enhancing Multimodal Understanding with CLIP-Based Image-to-Text Transformation. In Proceedings of the 2023 6th International Conference on Big Data Technologies (ICBDT '23). Association for Computing Machinery, New York, NY, USA, 414–418. https://doi.org/10.1145/3627377.3627442
[8] Lin, Q., Che, C., Hu, H., Zhao, X., & Li, S. (2023). A Comprehensive Study on Early Alzheimer's Disease Detection through Advanced Machine Learning Techniques on MRI Data. Academic Journal of Science and Technology, 8(1), 281–285.DOI: 10.1111/jgs.18617
[9] Che, C., Hu, H., Zhao, X., Li, S., & Lin, Q. (2023). Advancing Cancer Document Classification with R andom Forest. Academic Journal of Science and Technology, 8(1), 278–280. https://doi.org/10.54097/ajst.v8i1.14333
[10] S. Tianbo, H. Weijun, C. Jiangfeng, L. Weijia, Y. Quan and H. Kun, "Bio-inspired Swarm Intelligence: a Flocking Project With Group Object Recognition," 2023 3rd International Conference on Consumer Electronics and Computer Engineering (ICCECE), Guangzhou, China, 2023, pp. 834-837, doi: 10.1109/ICCECE58074.2023.10135464.
[11] Y. Wang, K. Yang, W. Wan, Y. Zhang and Q. Liu, "Energy-Efficient Data and Energy Integrated Management Strategy for IoT Devices Based on RF Energy Harvesting," in *IEEE Internet of Things Journal*, vol. 8, no. 17, pp. 13640-13651, 1 Sept.1, 2021, doi: 10.1109/JIOT.2021.3068040.
[12] Y. Wang, K. Yang, W. Wan, Y. Zhang and Q. Liu, "Energy-Efficient Data and Energy Integrated Management Strategy for IoT Devices Based on RF Energy Harvesting," in IEEE Internet of Things Journal, vol. 8, no. 17, pp. 13640-13651, 1 Sept.1, 2021, DOI: 10.1109/JIOT.2021.3068040.
[13] Xu, J., Pan, L., Zeng, Q., Sun, W., & Wan, W. Based on TPUGRAPHS Predicting Model Runtimes Using Graph Neural Networks. https://api.semanticscholar.org/Corpus
[14] Yao, J., Zou, Y., Du, S., Wu, H., & Yuan, B. Progress in the Application of Artificial Intelligence in Ultrasound Diagnosis of Breast Cancer. DOI:https://api.semanticscholar.org/Corpus
[15] Zhou Y, Chen S, Wu Y, Li L, Lou Q, Chen Y, Xu S. Multi-clinical index classifier combined with AI algorithm model to predict the prognosis of gallbladder cancer. Front Oncol. 2023 May 10;13:1171837. DOI: 10.3389/fonc.2023.1171837. PMID: 37234992; PMCID: PMC10206143.

[16] Li L, Xu C, Wu W, et al. Zero-resource knowledge-grounded dialogue generation[J]. Advances in Neural Information Processing Systems, 2020, 33: 8475-8485. DOI: https://doi.org/10.48550/arXiv.2008.12918